\documentclass{DISproc}
\def\tb{\bar t}
\def\ttb{{t\tb}}
\newcommand\sss{\mathchoice%
{\displaystyle}%
{\scriptstyle}%
{\scriptscriptstyle}%
{\scriptscriptstyle}%
}
\newcommand\MCatNLO{{\tt MC@NLO}}
\newcommand\POWHEG{{\tt POWHEG}}
\newcommand\POWHEGBOX{{\tt POWHEG BOX}}
\newcommand\HERWIG{{\tt HERWIG}}

\newcommand\PYTHIA{{\tt PYTHIA}}

\newcommand\pt{p_{\sss\rm T}}
\begin{document}
\title{
Top-quark pair-production with one jet and parton showering at hadron colliders}

\author{{\slshape Simone Alioli$^1$, Juan Fuster$^2$, Adrian Irles$^2$,
    Sven-Olaf Moch$^3$, Peter Uwer$^4$, Marcel Vos$^2$}\\[1ex]
$^1$LBNL \& UC Berkeley, 1 Cyclotron Road, Berkeley, CA 94720, USA\\
$^3$IFIC, Universitat de Valencia -- CSIC, Catedratico Jose Beltran 2, E-46980 Paterna, Spain\\
$^3$DESY, Platanenalle 6, D-15738 Zeuthen, Germany\\
$^4$Humboldt-Universit{\"a}t, Newtonstra{\ss}e 15, D-12489 Berlin, Germany }

\contribID{xy}

\doi  

\maketitle

\begin{abstract}
We present heavy-flavor production in association with one jet 
in hadronic collisions matched to parton shower Monte Carlo predictions 
at next-to-leading order QCD with account of top-quark decays 
and spin correlations. 
We use the \POWHEGBOX{} for the interface to the parton shower programs \PYTHIA{} or \HERWIG{}. 
Phenomenological studies for the LHC and the Tevatron are presented with particular 
emphasis on the inclusion of spin-correlation effects in top decay 
and the impact of the parton shower on the top-quark charge asymmetries. 
As a novel application of the present calculation the measurement 
of the top-quark mass is discussed.
\end{abstract}

\vspace*{-1mm}
\section{Introduction}
\vspace*{-2mm}
The Large Hadron Collider (LHC) and the Tevatron provide an experimental environment 
allowing for top-quark measurements with percent level accuracy.
Precise measurements for top-quark production demand theoretical
predictions with comparable precision.  This requires the knowledge of the hard scattering process
beyond the leading order (LO) in perturbation theory. Furthermore, for the direct comparison with experimental data, fully exclusive events are needed, that take into account all-order logarithmic enhancements of soft and collinear regions of phase space  and hadronization effects by means of Shower Monte Carlo (SMC) programs.
Both approaches can be combined systematically by merging NLO computations with parton
showers, in the \MCatNLO{}~\cite{Frixione:2002ik} or \POWHEG{}~\cite{Nason:2004rx} approach.

\vspace*{-1mm}
\section{$\boldsymbol{\ttb} + \mbox{1-jet}$ hadroproduction  in \POWHEG{}} 
\vspace*{-2mm}
In the following we concentrate on the recent implementation of the
$\ttb + \mbox{1-jet}$ hadroproduction  in the \POWHEG{} approach,  presented in Ref.~\cite{Alioli:2011as}.  
A large fraction of
the inclusive $\ttb$ production does  indeed actually contain events with one or even more
additional jets. Furthermore, due to the larger phase space available, the relative
importance of data samples with $\ttb +$jets is larger at the LHC with
respect to the Tevatron, increasing the need of an accurate theoretical description of this process.
  Top-quark pair-production associated with
jets is also an important background to Higgs boson production in
vector boson fusion and for many signals of new physics.  The
implementation reported here is based on the NLO QCD corrections evaluated
in Ref.~\cite{Dittmaier:2007wz,Dittmaier:2008uj}, merged with
\HERWIG{}~\cite{Corcella:2000bw} and \PYTHIA{}~\cite{Sjostrand:2006za}
SMC programs, using the \POWHEGBOX{}~\cite{Alioli:2010xd}.  

We present results for both Tevatron and LHC colliders, having assumed
a jet reconstruction cut in the analysis of $\pt > 20$~GeV and
$50$~GeV, respectively.  We have used the inclusive-$k_T$ jet algorithm with
$R = 1$ and the $E_T$-recombination scheme. Renormalization and
factorization scales have been set to $\mu_R = \mu_F = m_t = 174~$GeV,
we have used the PDF set CTEQ6M~\cite{Pumplin:2002vw}, and we have not imposed any extra acceptance
cut, other than those necessary to define the hard jet.
\begin{figure}[t]
\begin{minipage}[b]{0.475\linewidth}
\centering
\includegraphics[width=0.95\textwidth]{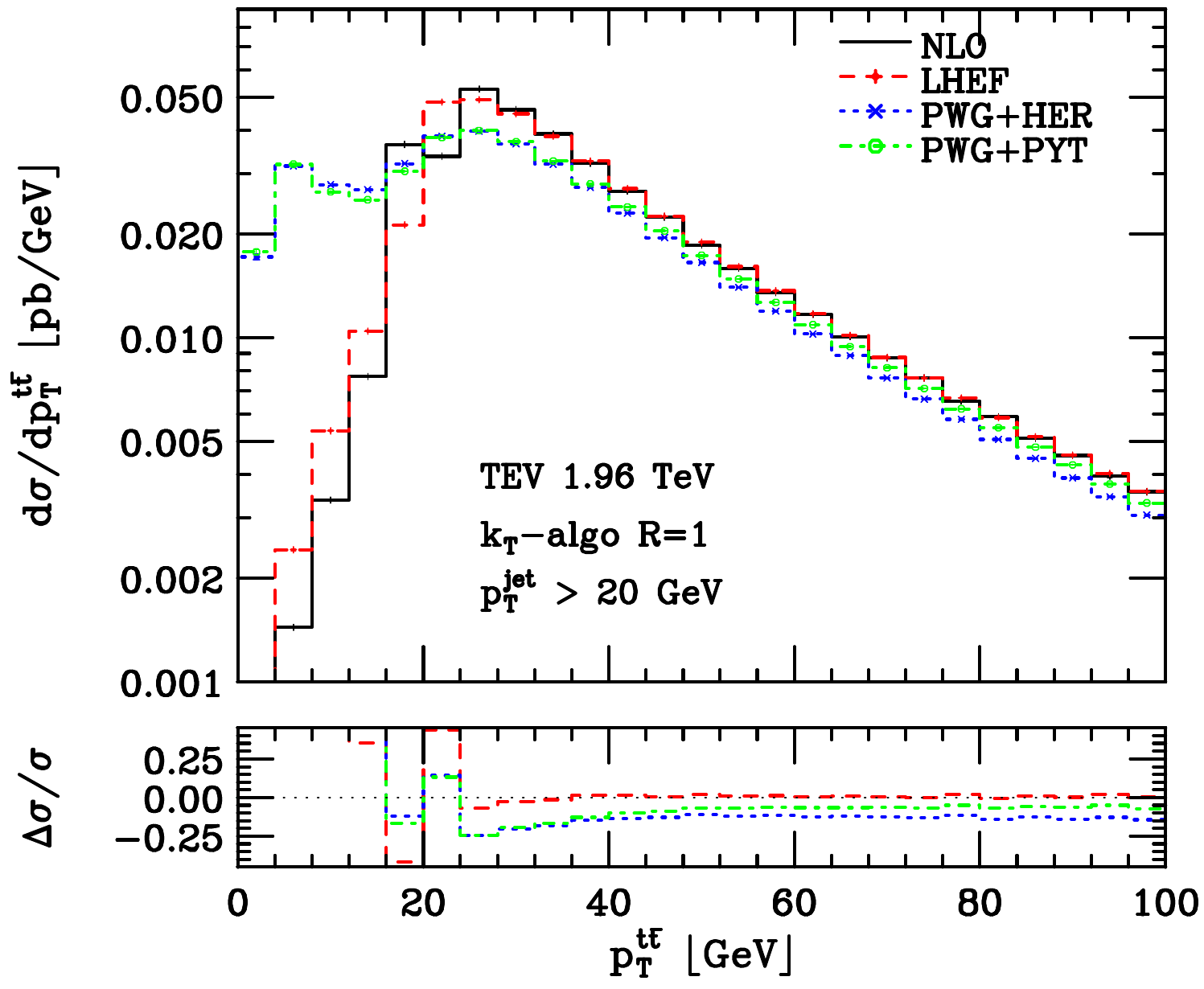}
\caption{Differential cross section as a function of the $\ttb$-pair transverse momentum at the  Tevatron ($\sqrt s =1.96$~TeV)}
\label{Fig:1}
\end{minipage}
\hspace{20pt}
\begin{minipage}[b]{0.475\linewidth}
\centering
\includegraphics[width=0.95\textwidth]{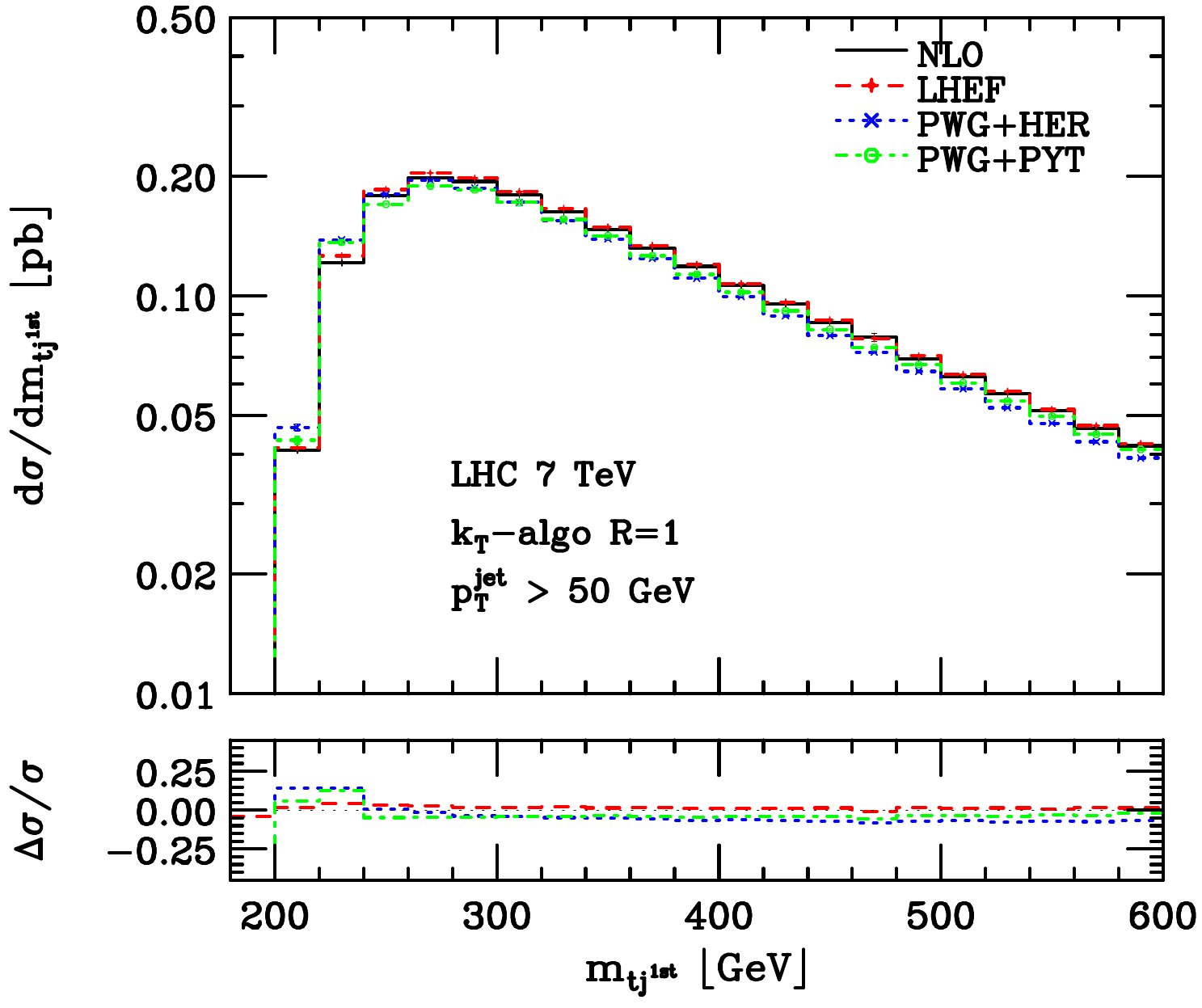}
\caption{Differential cross section as a function of the $(t j_1)$ invariant mass at the LHC ($\sqrt s =7$~TeV)}
\label{Fig:2}
\end{minipage}
\end{figure}
In Fig.~\ref{Fig:1} we show the differential cross section as a function of the transverse momentum of the $\ttb$-pair at the Tevatron, while in Fig.~\ref{Fig:2} we plot the invariant mass of the system made by the top-quark and the hardest jet at the $7$~TeV LHC. The different curves appearing on each plot refer respectively to the fixed order results ({\sc nlo}), to the results after the first emission has been performed by \POWHEG{} ({\sc lhef}) and to the fully showered events, with \HERWIG{} ({\sc pwg+her}) or \PYTHIA{} ({\sc pwg+pyt}) showers. Shower effects are visible in the low-$\pt^{\ttb}$ region, while more inclusive observables like the invariant mass of the system made by the top-quark and the hardest jet, $m_{(t j_1)}$, are basically unaffected by the shower.

\vspace*{-1mm}
\section{Spin correlations in top-quark decays}
\begin{figure}[t]
\begin{minipage}[b]{0.475\linewidth}
\centering
\includegraphics[width=0.95\textwidth]{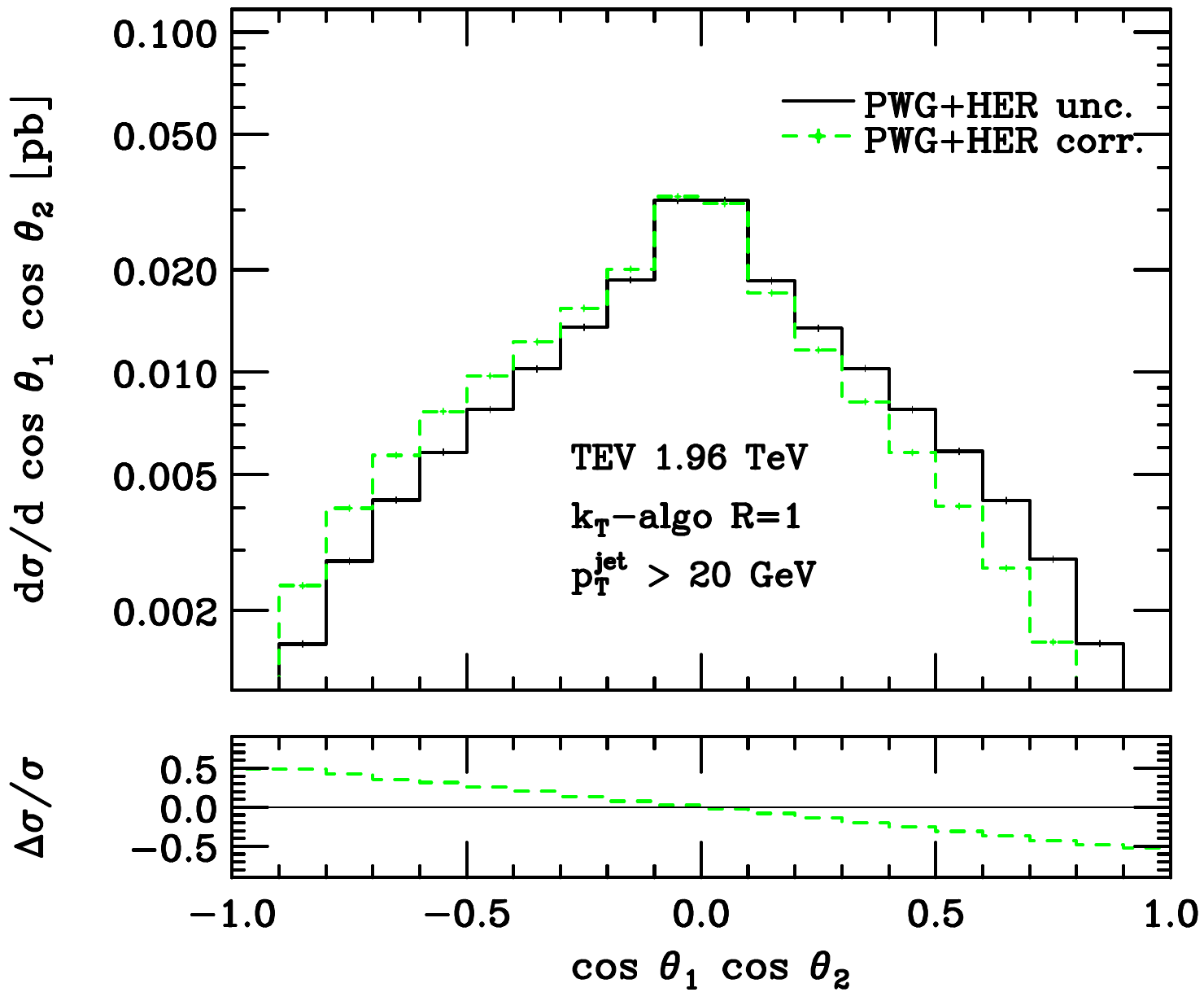}
\caption{Effect of the inclusion of spin correlations  when interfacing to \HERWIG{}.}
\label{Fig:3}
\end{minipage}
\hspace{20pt}
\begin{minipage}[b]{0.475\linewidth}
\centering
\includegraphics[width=0.95\textwidth]{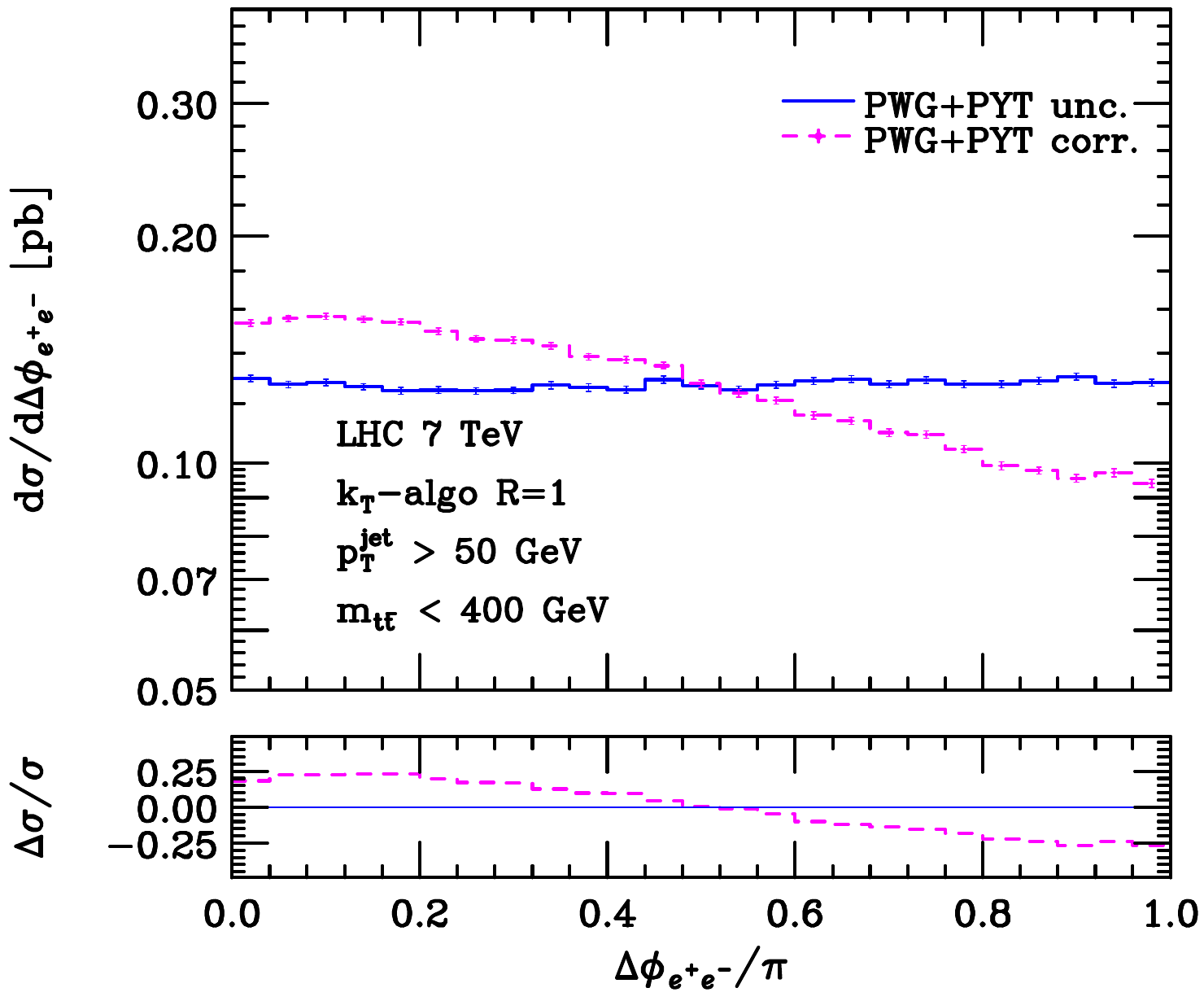}
\caption{Effect of the inclusion of spin correlations  when interfacing to \PYTHIA{}.}
\label{Fig:4}
\end{minipage}
\end{figure}

\vspace*{-2mm}
In our implementation we have also included the spin-correlations. 
In doing so, we have neglected off-shell effects and non-resonant production mechanisms. We
proceeded by first generating events with stable top-quarks
(un-decayed events) through the usual \POWHEG{} machinery and then
generating the decay products according to the matrix element for the
full production and decay process (decayed events),
following Ref.~\cite{Frixione:2007zp}.  In our study we always assumed the
double-leptonic top-quark decay channel $t\to W^+ b \to \ell^+ \nu b$. 
In Fig.~\ref{Fig:3} we draw the differential distribution
$\frac{1}{\sigma} \frac{d^2\sigma}{d\cos \theta_1 d\cos \theta_2}$ after the \HERWIG{} shower, at
the Tevatron collider, where the angles $\theta_1$ and $\theta_2$ 
between the directions of flight of the leptons coming from the
decayed top-quark in the $t$ ($\tb $) rest frame and the beam axis 
can be interpreted in the context of spin correlations as the quantization
axis for the \mbox{(anti-)}top-quark spin. 
No extra acceptance cut is imposed on the leptons.  In
Fig.~\ref{Fig:4} we show instead the differential cross section as a function of the azimuthal distance between the
two leptons coming from the top-quarks decays, for the LHC collider
configuration and after the \PYTHIA{} shower. 
An extra cut $m_{\ttb}<400$~GeV
has been imposed here to enhance the effect.  
A similar observable has recently been used in $t\bar{t}$-production to
observe spin-correlations~\cite{ATLAS:2012ao}.
The plots in Fig.~\ref{Fig:4} clearly demonstrate the differences between spin-correlated results 
and those obtained by letting the respective SMC program performing uncorrelated top-quark decays.

\vspace*{-1mm}
\section{Asymmetries}
\vspace*{-2mm}
We have also investigated the $\ttb$ charge  asymmetry in presence of a hard
jet, finding that the inclusion of the parton shower changes significantly
the fixed-order predictions in the low $p_T^{\ttb}$ region, where
shower effects are known to be large.  
Away from this region the
parton shower leads only to a marginal change of the charge asymmetry
binned in $p_T^{\ttb}$. 
This quantity is now available at NLO accuracy, supplemented by the shower. 
For more details and for complete tables including results obtained 
with different cuts and at various stages of the simulation, 
we refer to Ref.~\cite{Alioli:2011as}.

\vspace*{-1mm}
\section{Top-quark mass measurement}
\begin{figure}[t]
\begin{minipage}[b]{0.475\linewidth}
\centering
\includegraphics[width=1.0\textwidth]{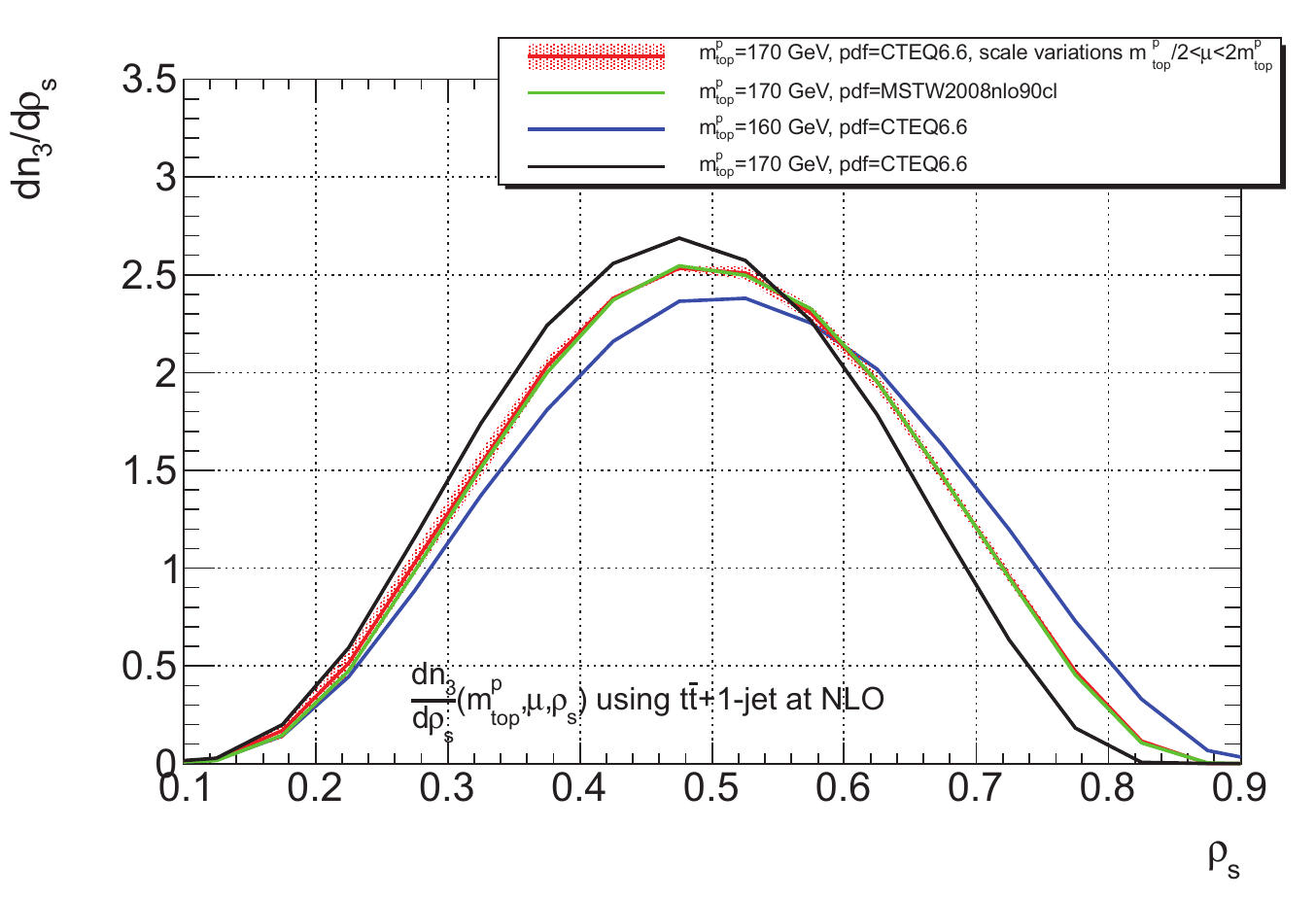}
\caption{$dn_{3}/d\rho_{s}(m_{top}^{p},\mu)$
  calculated at NLO for different masses $m_{top}^{p}=160,\,170$ and
  $180$\, GeV. For $m_{top}^{p}=170$\,GeV the scale uncertainty is shown 
  and two PDF sets~\cite{Nadolsky:2008zw,Martin:2009iq} for comparison.}
\label{Fig:5}
\end{minipage}
\hspace{20pt}
\begin{minipage}[b]{0.475\linewidth}
\centering
\includegraphics[width=1.0\textwidth]{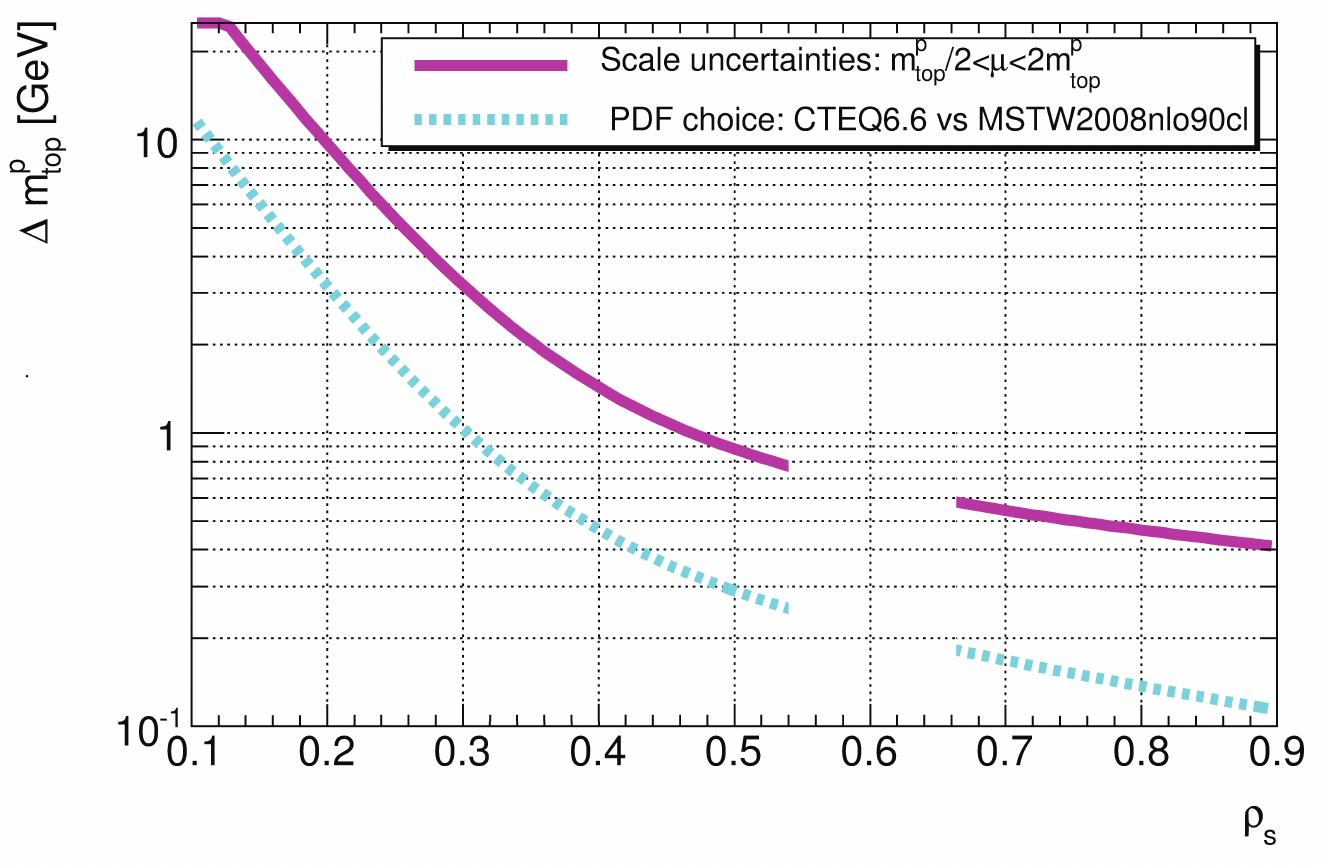}
\caption{Sensitivity on value of 
  the top-quark mass (for $m_{top}^{p}=170\,\rm GeV$)
  along with scale uncertainty (magenta solid line) 
  and effect of PDF choice~\cite{Nadolsky:2008zw,Martin:2009iq}. 
  The crossing region is excluded.}
\label{Fig:6}
\end{minipage}
\end{figure}

\vspace*{-2mm}
As a novel application the differential cross section for $t\bar{t}+\mbox{1-jet}$ production
can be used for a determination of the top-quark mass.
To that end, we consider the differential $t\bar{t}+\mbox{1-jet}$ rate,
\begin{eqnarray}
\label{dist1}
\frac{dn_{3}}{d\rho_{s}}(m_{top}^{p},\mu,\rho_{s})=
\frac{1}{\sigma_{t\bar{t}j}} 
\frac{d\sigma_{t\bar{t}j}}{d\rho_{s}}(m_{top}^{p},\mu,\rho_{s}),
\end{eqnarray}
where $\sigma_{t\bar{t}j}$ denotes the cross section for the process $pp\to t\bar{t}+\mbox{1-jet} + X$. 
The variable $\rho_{s}$ is defined as $\rho_{s}=\frac{2\cdot m_{0}}
{\sqrt{s_{t\bar{t}j}}}$ with $m_{0}=170$\, GeV 
and $s_{t\bar{t}j}$ is the invariant mass squared of the final state. 
In Fig.~\ref{Fig:5} a clear separation between the distributions for different top-quark masses is observed 
except in the region of $0.55<\rho_{s}<0.62$ where the curves cross due to the normalization
of $\frac{dn_{3}}{d\rho_{s}}(m_{top}^{p},\mu,\rho_{s})$. 
As a consequence a decrease of sensitivity is observed in the crossing region.
The approach of Eq.~(\ref{dist1}) nicely complements top-quark mass
measurements from the $t\bar{t}$ total cross section, see
Ref.~\cite{Langenfeld:2009wd} for the first measurement of the
$\overline{\mbox{MS}}$ mass.
The impact of the conventionally estimated scale variation (solid line) 
and the PDF choice (dashed line) on the top mass value (for $m_{top}=170$\,GeV) 
is displayed in Fig.~\ref{Fig:6}. 
It demonstrates that a theoretical uncertainty of $500-600\,$ MeV can be reached 
with a mass measurement in the interval $\rho_{s}>0.62$ based on the scale 
uncertainty and the dependence on the PDF choice.
Additional sources of systematic uncertainities have been investigated and have led to 
error estimates below $1\,$ GeV.
The crossing region is again excluded due to the vanishing sensitivity. 
The curves in Fig.~\ref{Fig:6} have been obtained assuming a linear dependence of $n_3$ on the 
top-quark mass for intervals of $\Delta m_{top}^{p}=5\,$ GeV.


{\raggedright
\begin{footnotesize}

\providecommand{\href}[2]{#2}\begingroup\raggedright\endgroup

\end{footnotesize}
}


\end{document}